\documentclass[multphys,vecphys]{svmult}

\usepackage{makeidx}         
\usepackage{graphicx}        
\usepackage{multicol}        
\usepackage[bottom]{footmisc}

\makeindex             

\begin{document}

\title*{Ultra-Compact Stellar Systems in the Fornax Galaxy Cluster}
\author{P. Firth\inst{1}\and
M.J. Drinkwater\inst{1}\and
E.A. Evstigneeva\inst{1}\and
A. Karick\inst{2}\and
M.D. Gregg\inst{2}\and
M. Hilker\inst{3}\and
K. Bekki\inst{4}\and
J.B. Jones\inst{5}\and
S. Phillipps\inst{6}}
\institute{Dept. of Physics, University of Queensland, Qld 4072, Australia
\texttt{firth@physics.uq.edu.au}
\and University of California, Davis
\and University of Bonn
\and School of Physics, University of New South Wales
\and Queen Mary, University of London
\and University of Bristol}
%
%
\titlerunning{UCDs in the Fornax Galaxy Cluster}
\authorrunning{Firth et al.}
\maketitle


\section{Observing the UCD/GC Interface}
\label{sec:1}
Ultra-compact dwarfs (UCDs) are massive but compact gravitationally-bound stellar systems discovered in the nearby Fornax \cite{Hilker1999}\cite{Drinkwater2000}\cite{Drinkwater2004}\cite{Mieske2002}\cite{Phillipps2001} and Virgo \cite{Jones2006} galaxy clusters. Several UCD formation theories have emerged -– that they are ultra-massive examples of globular clusters (GCs) \cite{Mieske2002}; or stellar super-clusters created in gas-rich galaxy mergers \cite{Fellhauer2002}; or the remnant cores of tidally-stripped nucleated dwarf galaxies \cite{Bekki2001}.

We completed spectroscopic observations in November 2004 of colour-selected point source targets (18.00 \textless r{\footnotesize{$^\prime$}} \textless 22.75) in four 25{\footnotesize$^\prime$} diameter VLT fields surrounding NGC1399, the massive cD galaxy at the core of the Fornax galaxy cluster (Figure 1:{\sc left}). Targets were selected from g{\footnotesize{$^\prime$}}r{\footnotesize{$^\prime$}}i{\footnotesize{$^\prime$}} imaging with the CTIO Blanco 4m telescope \cite{Karick2005}. We have discovered 30 new compact stellar systems at the cluster redshift, adding to 62 previously catalogued UCDs \cite{Hilker1999}\cite{Drinkwater2000}\cite{Drinkwater2004}. Our observations extend to the absolute magnitude range of globular clusters, enabling us to explore the UCD/GC interface ($-$12 \textless $M_{r'}$ \textless $-$9) in a forthcoming paper.

\section{UCD Radial Distribution and Kinematics}
\label{sec:2}
In Figure 1:{\sc right} the radial distribution of 51 previously known Fornax UCDs for which we have u{\footnotesize{$^\prime$}}g{\footnotesize{$^\prime$}}r{\footnotesize{$^\prime$}}i{\footnotesize{$^\prime$}}z{\footnotesize{$^\prime$}} photometry, and 30 new compact stellar systems from our VLT observations, is plotted against $M_{r'}$ magnitude. The new data show that compact stellar systems over a range of magnitudes are found extensively in intra-cluster space.

The UCD system has a mean velocity of $1478 \; \mbox{km} \, \mbox{s}^{-1}$ and a dispersion ($\sigma${\footnotesize{$_0$}}) of $244 \; \mbox{km} \, \mbox{s}^{-1}$. Our redshift data show weak evidence for a net rotation of the 92-member UCD system about NGC1399. The velocity gradient is $57 \pm 42 \; \mbox{km} \,\mbox{s}^{-1} \, \mbox{deg}^{-1}$ in R.A. from NGC1399 (V/$\sigma${\footnotesize{$_0$}} = 0.23 $\pm$ 0.17). This rotation contrasts with the finding \cite{Richtler2004} that the inner GC population (2{\footnotesize{$^\prime$}} to 9{\footnotesize{$^\prime$}} radius) shows little or no rotation about NGC1399.

\begin{figure}
\centering
\includegraphics[height=5.5cm]{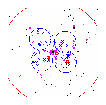}
\hspace{0.1 cm}
\includegraphics[height=5.5cm]{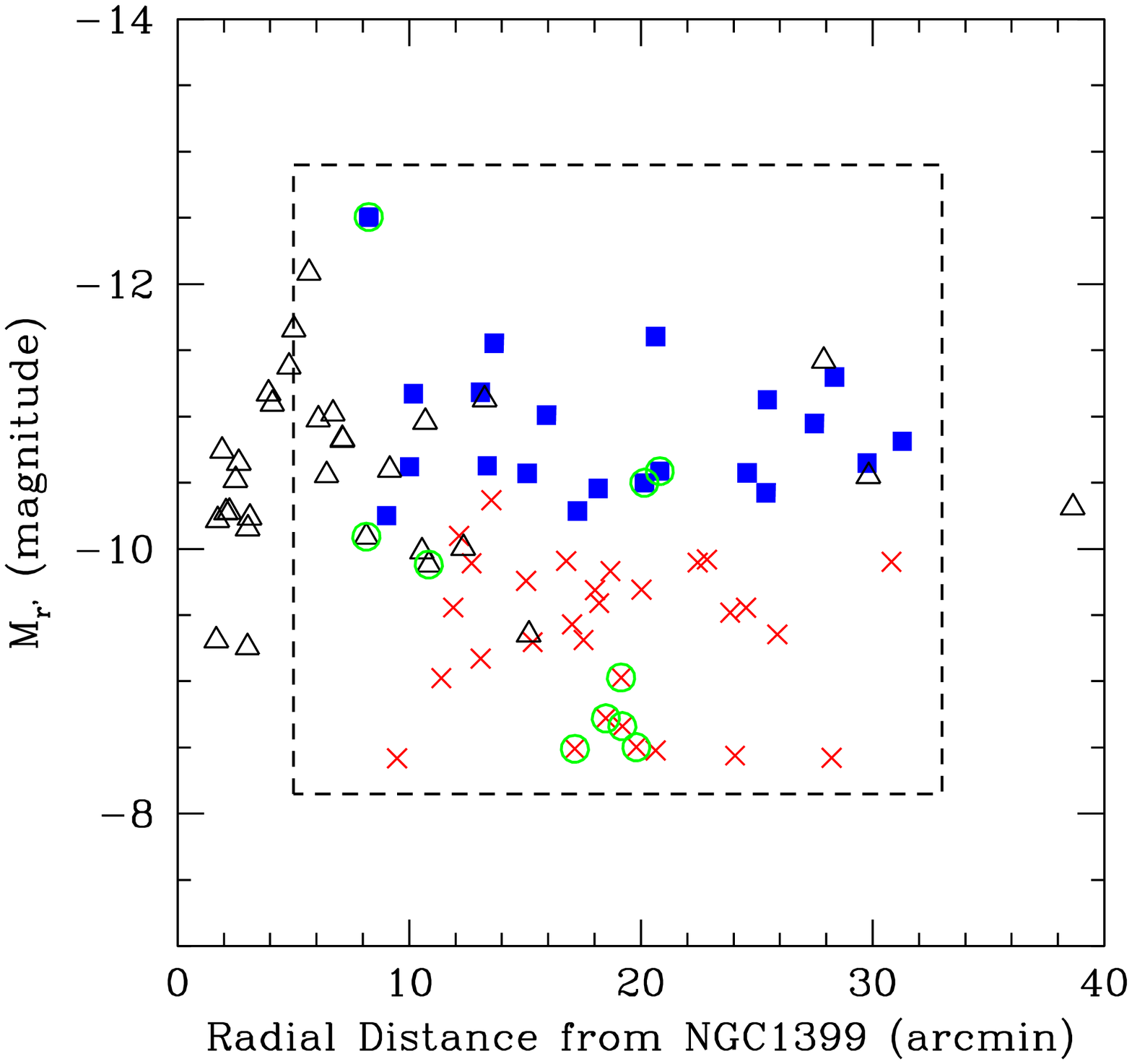}
\caption{\textbf{{\sc left}:} Approximately 1.5$^{\circ}$ square image of UCDs surrounding NGC1399. The 2dF field and four smaller VLT fields shown here contain previously catalogued UCDs (open small circles/squares) and newly discovered stellar systems (crosses). Our results suggest a UCD bar/filament structure stretches across NGC1399. \textbf{{\sc right}:} $M_{r'}$  magnitude against radial distance from NGC1399 (m $-$ M = 30.9) of 51 previously known UCDs, being those observed only with 2dF (triangles) or re-observed with VLT (squares), together with 30 newly-discovered UCDs (crosses). UCDs close to prominent cluster member galaxies other than NGC1399 are circled. The dashed rectangle shows the VLT field/magnitude outer limits.}
\label{fig:1}
\end{figure}

%
%
%
%
%
%

%
%



\printindex
\end{document}